# LCLS-II Technical Note

Wakefield Effects of the Bypass Line
in LCLS-II

LCLS-II TN-14-09

11/6/2014

K. Bane and T. Raubenheimer
SLAC, Menlo Park, CA 94025, USA

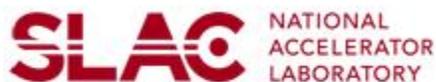
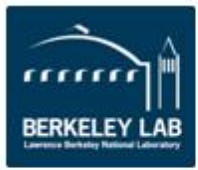
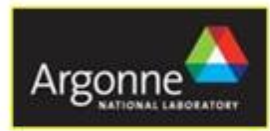
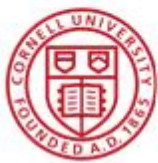

# WAKEFIELD EFFECTS OF THE BYPASS LINE IN LCLS-II[*]

K.L.F. Bane[§], T. Raubenheimer, SLAC, Menlo Park, CA 94025, USA

## INTRODUCTION

In LCLS-II [1], after acceleration and compression and just before entering the undulator, the beam passes through roughly 2.5 km of 24.5 mm (radius) stainless steel (SS) pipe. The bunch that passes through the pipe is extremely short with an rms of 8 $\mu$m for the nominal 100 pC case. Thus, even though the pipe has a large aperture, the wake that applies is the short-range resistive wall wakefield.

The LCLS-II bunch distribution is approximately uniform and therefore, for short bunches, the wake induced voltage is characterized by a rather linear variation. For self-seeding operation, however, longer bunches are desirable. An important question is, what constraints do the bypass line wakes place on the bunch length? In this note we calculate the wake, discuss the confidence in the calculation, and investigate how to improve the induced chirp linearity and/or strength. Finally, we also study the strength and effects of the transverse (dipole) resistive wall wakefield. Selected beam and machine parameters are given in Table 1.

Table 1: Selected beam and machine properties in the bypass of LCLS-II that are used in our calculations. The longitudinal bunch distribution is approximately uniform. Note that the nominal bunch charge is 100 pC, also with $I = 1$ kA.

| Parameter name | Value | Unit |
| --- | --- | --- |
| Charge per bunch, $Q$ | 300 | pC |
| Beam current, $I$ | 1 | kA |
| Rms bunch length, $\sigma_z$ | 25 | $\mu$m |
| Beam energy, $E$ | 4 | GeV |
| Maximum repetition rate, $f_{rep}$ | 1 | MHz |
| Bypass pipe radius, $a$ | 2.45 | cm |
| Bypass pipe length, $L$ | 2.5 | km |
| Average $\beta$ function $\beta_x$ | 260 | m |

## WAKE OF BYPASS LINE

The point charge (dc) resistive wall wake in a round, metallic pipe is given by [2]

$$W_\delta(s) = \frac{Z_0 c}{\pi a^2}\left(\frac{1}{3}e^{-s/s_0}\cos\frac{\sqrt{3}s}{s_0} - \frac{\sqrt{2}}{\pi}\int_0^\infty dx\,\frac{x^2 e^{-sx^2/s_0}}{x^6+8}\right) \quad (1)$$

with $s$ ($>0$) the distance the test particle follows behind the drive particle, $Z_0 = 377\,\Omega$, $c$ the speed of light, and $a$ the beam pipe radius. The characteristic distance over which the wake falls to zero, $s_0 = (2a^2/Z_0\sigma_c)^{1/3}$, with $\sigma_c$ the conductivity of the walls.

[*] Work supported by Department of Energy contract DE–AC02–76SF00515.
[§] kbane@slac.stanford.edu

For the LCLS-II bypass line $a = 2.45$ cm and for SS $\sigma_c = 1.4\times 10^6\,\Omega^{-1}\text{m}^{-1}$. Thus $s_0 = 132\,\mu$m. A plot of the total wake of the bypass is given in Fig. 1.

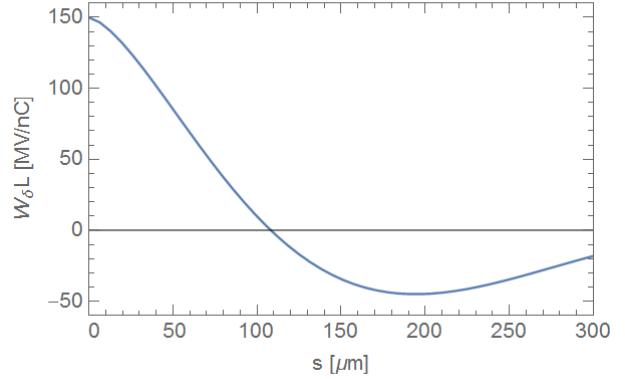

Figure 1: Point charge wake of LCLS-II bypass beam pipe.

The bunch shape in the LCLS-II bypass line is approximately uniform. For a uniform bunch distribution the induced relative energy variation is given by

$$\delta_w(s) = -\frac{eIL}{cE}\int_0^s W(s')\,ds' \qquad 0 \le s \le \ell, \quad (2)$$

with $I$ the beam current, $L$ the pipe length, $E$ the beam energy, with the head of the bunch at $s = 0$ and the tail at $s = \ell$. The induced energy variation for the bypass line, assuming a uniform bunch with $I = 1$ kA, is plotted in Fig. 2. The head of the bunch is at $s = 0$, and the extent of three charge scenarios is indicated by color coding. The bunch reaches to $\ell = 2\sqrt{3}\sigma_z = Qc/I = 30, 90, 150\,\mu$m for the cases $Q = 100, 300, 500$ pC, respectively. We note that $\delta_w$, for the 100 pC case, introduces a linear chirp of $\sim -10^{-4}\,\mu\text{m}^{-1}$, for the 300 pC case the chirp has curvature in the bunch tail, and for the 500 pC case the chirp at the tail actually changes sign.

A linear induced chirp is desirable, since it is of the correct sign to help remove surplus chirp left over from the last bunch compressor (to "dechirp" the beam). But the curvature at the tail of the 500 pC case is not desirable, since it will leave a non-correctable energy variation in the tail of the bunch. Because of this result, the 500 pC scenario has been dropped as a mode of operation for the LCLS-II.

How could one improve the situation for the 500 pC case? To have the chirp be rather linear for this case, one could *e.g.* increase $s_0$ by coating the inside of the bypass pipe with a metal that is a poorer conductor than SS. For example, to have the curvature of the chirp be the same or better than the 300 pC case with SS, the coating metal would need to be at least $(5/3)^3 = 5$ times worse than SS, such as can be obtained with a layer of graphite.

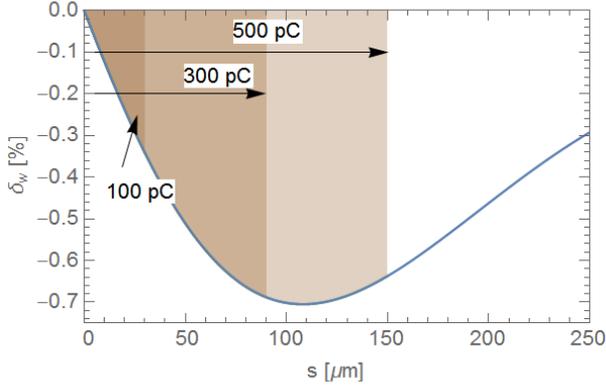

Figure 2: Relative wakefield-induced energy change in the bypass, for a uniform bunch distribution with $I = 1$ kA. The head of the bunch is at $s = 0$, and the extent of three charge scenarios is indicated by color coding.

*Actual Bunch Shapes*

For bunches of general shape, the induced energy variation is given by the convolution

$$\delta_w(s) = -\frac{eL}{cE} \int_0^\infty W(s') I(s - s') \, ds' \ . \quad (3)$$

In Fig. 3 we plot bunch shapes and induced relative energy variation for the cases $Q = 100$ pC (top) and 300 pC (bottom). The bunch shapes were obtained by numerical simulations with the computer program LiTrack [3]. The dashed curves give the result assuming a uniform distribution with $I = 1$ kA. In spite of small horns at the head of the bunch, $\delta_w(s)$ for the more accurate calculation agrees quite well with that using the uniform assumption (at least over the bunch core).

## CONFIDENCE IN WAKE

The wake of the bypass line impacts significantly the longitudinal phase space of the LCLS-II beam. How confident are we that the calculation is correct for the bypass line? Here are supporting arguments:

- We note that there are only two parameters in the wake, the wake amplitude and the longitudinal scale factor $s_0$, which depend only on the physical parameters $a$ and $\sigma_c$. And the dependence on conductivity ($s_0 \sim \sigma_c^{-1/3}$) is weak and insensitive to inaccuracies in $\sigma_c$.

- A more accurate formulation of the resistive wall wake, the *ac* resistive wall wake assumes that the effective wall conductivity varies with frequency as $\tilde{\sigma}_c(\omega) = \sigma_c/(1 - i\omega\tau)$, with $\tau$ the relaxation time of the metal [2].

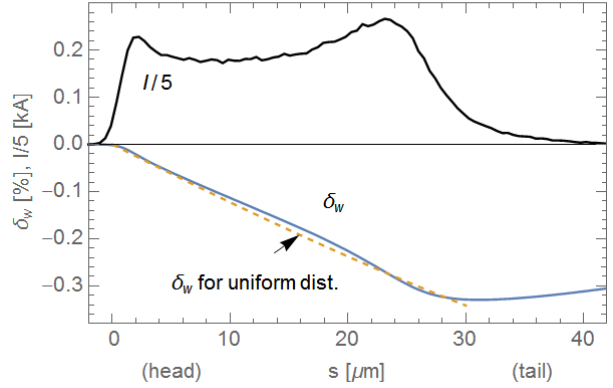

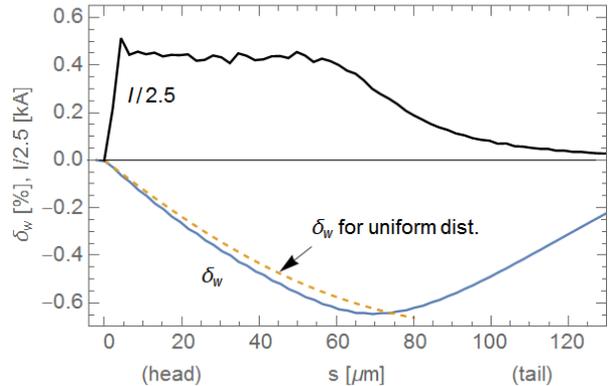

Figure 3: Relative induced voltage for realistic bunch distributions (blue), for $Q = 100$ pC (top), 300 pC (bottom). Dashes gives result for uniform distributions. Current shape given in black, with head to the left. Bunch shapes were obtained by L. Wang.

However, the corresponding wake formula gives close to the same result as the dc result (Eq. 1), whenever the ratio $c\tau/s_0$ is small compared to 1. Here this ratio is very small because $\tau$ is small for SS and because $s_0$ here is relatively large.

- When the beam first enters the bypass line it excites a trainsient wake. After a distance on the order of the so-called catch-up distance, $z_{cu} = a^2/2\sigma_z$, the wake reaches steady-state, where the formula of Eq. 1 applies. In LCLS-II, for the nominal 100 pC case, $\sigma_z = 8$ $\mu$m and the catch-up distance $z_{cu} = 58$ m. This length is very short compared to the 2.5 km length of the bypass; consequently, the transient contribution can be ignored.

- The normal wakefield formulas are valid provided the bunch length $\sigma_z \gtrsim a/\gamma$. At wave numbers beyond $\gamma/a$ the impedance rapidly drops to zero. At 4 GeV, $a/\gamma = 3$ $\mu$m, which is still smaller than $\sigma_z = 8$ $\mu$m in the 100 pC case, and the wake formulas apply.

- We know how to estimate the wake effects of surface roughness; here, however, $a$ and the sensitivity to roughnes must be small. At high frequency the anomalous skin effect (ASE) can change the rw wake; however, ASE is primarily a low temperature effect, and also SS doesn't seem to exhibit ASE.

With these arguments we believe that the wake formula, Eq. 1, is applicable and accurate for LCLS-II bunches with $I = 1$ kA and $Q \gtrsim 100$ pC, the nominal bunch charge. That having been said, there has never been a direct measurement of the short-range resistive wall wakefield. Energy loss (not induced chirp) measurements were performed in the undulator region—where the short-range rw wake is also thought to dominate—in LCLS-I in 2009 [4]. Those measurements agreed quite well with calculations [5–7], though there were complications to the measurement that added some uncertainty to this conclusion.[1]

## DIPOLE WAKE

Since the LCLS-II bunch length is extremely small we expect the effect of the transverse (dipole) wake to be small. Nevertheless, for completeness, we also consider the dipole wake effect of the LCLS-II bypass line. The point charge, resistive wall dipole wake is calculated in a similar manner, using an equation similar to Eq. 1 [2]. It is plotted in Fig. 4 (the blue curve). Note that he slope at the origin is given by $W'_x(0^+) = 2Z_0 c/(\pi a^4)$ (shown by the dashed line). The dipole wake peaks near $s = s_0$.

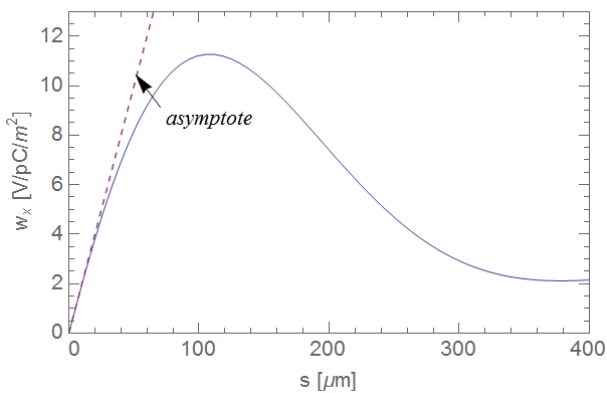

Figure 4: Dipole point charge wake of LCLS-II bypass beam pipe. The slope of the wake at the origin is given by the dashed line.

---

[1] Only the combined effect of the Linac-To-Undulator (LTU) transfer line and the 100-m undulator vacuum chamber could be measured. The LTU chamber is rather complicated and was estimated to contribute $\sim 20\%$ to the measured loss.

The single bunch, beam break-up (BBU) can be characterized by the strength parameter $\Upsilon = eQ\beta_x W_{\lambda,x}(s)/2E$, with $\beta_x$ the average beta function and $W_{\lambda,x}$ the *bunch* dipole wake [8]. If $\Upsilon$ is small compared to 1, then the injection jitter amplification is small and the bypass line misalignment tolerance is relaxed and similar to the zero current one.

For the 300 pC case, taking $\beta_x = 260$ m, we find that at the tail of the bunch, $\Upsilon(l) = 0.2$. The transverse wake effects are indeed weak.

## CONCLUSION

The wakefield of the bypass line of LCLS-II will add a fairly linear relative energy chirp of $10^{-4}/\mu$m to the beam for the nominal 100 pC case and also for the high charge 300 pC case (assuming a peak current of $I = 1$ kA and energy $E = 4$ GeV). This will help in removing residual energy chirp left in the beam from the final bunch compression. A 500 pC scenario (also with 1 kA), however, will induce a chirp with the wrong sign in the tail of the bunch, and is not acceptable.

Since the wake only depends on beam pipe radius $a$ and wall metal conductivity $\sigma_c$, and is also insensitive to errors in the knowledge of the latter, we can have confidence in the calculated result. Finally, since the bunch is short, even for the 300 pC case, the transverse wake effects are negligible.

## ACKNOWLEDGEMENTS

The authors thank L. Wang for generating the bunch shapes presented in Fig. 3.

## REFERENCES

[1] LCLS-II Conceptual Design Report, LCLSII-1.1-DR-0001-R0, November 2013.

[2] K. Bane and M. Sands, AIP Conf. Proc. 367 (1996) 131.

[3] K. Bane and P. Emma, Proceeding of 2005 Particle Accelerator Conf. PAC05, Knoxville, TN, p. 4266.

[4] LCLS Log book, March 2009.

[5] Juhao Wu, talk given at SLAC, at LCLS meeting in March 2009.

[6] A. Novokhatski, Proceedings of 2009 FEL Conf., FEL09, Liverpool, UK, p. 550.

[7] K. Bane, talk given at SLAC, at LCLS-II meeting, on March 5, 2014.

[8] A. Chao, Physics of collective beam instabilities in high energy accelerators, (John Wiley & Sons, New York, 1993), Chap. 3.